\documentclass[epj]{svjour}
\usepackage{graphics}
\begin{document}
\title{Transverse Enhancement Model and MiniBooNE Charge Current Quasi-Elastic Neutrino Scattering Data}
\author{Jan T. Sobczyk
}                     
%
%
\institute{Institute of Theoretical Physics\\ Wroc\l aw University}
\date{Received: date / Revised version: date}
%
\abstract{
Recently proposed Transverse Enhancement Model of nuclear effects in Charge Current Quasi-Elastic neutrino scattering [A. Bodek, H. S. Budd, and M.
E. Christy,  Eur. Phys. J. C{\bf 71} (2011) 1726] is confronted with the MiniBooNE high statistics experimental data. It is shown that the {\it effective} large axial mass model leads to better agreement with the data.
\PACS{ {13.15.+g} {Neutrino interactions} \and {25.30.Pt} {Neutrino scattering}
     } 
} 
\authorrunning{J.T. Sobczyk}
\titlerunning{Transverse Enhancement Model and the MiniBooNE CCQE data}
\maketitle
\section{Introduction}
\label{intro}
Charge current quasi-elastic (CCQE) scattering is the most abundant neutrino 
interaction in oscillation experiments like MiniBooNE (MB) or T2K with a flux
spectrum peaked below $1$~GeV. Its full understanding
is important for detail neutrino oscillation pattern measurements. 

Under an assumption that the impulse approximation picture \cite{impulse} is valid the CCQE reaction both on free and bound nucleons is defined as: 
\begin{equation}
\nu + n \rightarrow l^- + p\qquad {\rm or}\qquad 
\bar \nu + p \rightarrow l^+ + n\label{definition}
\end{equation}
with $\nu$, $\bar\nu$, $l^\pm$, $p$ and $n$ standing for: neutrino, antineutrino, charged lepton, proton and neutron respectively.

A theoretical description of free nucleon target CCQE reaction 
is based on the conserved vector current (CVC) and the partially conserved 
axial current (PCAC) hypotheses. The
only unknown quantity is then the axial form-factor $G_A(Q^2)$ for 
which one typically 
asssumes the dipole form $G_A(Q^2)=G_A(0)(1+\frac{Q^2}{M_A^2})^{-2}$ with 
a free parameter, called the axial mass $M_A$. 

The aim of CCQE cross section measurements is to determine the value 
of $M_A$ and also to validate a nuclear physics input used in theoretical cross section computations. There is a variety of approaches \cite{theories} including the Fermi Gas (FG) model implemented in all the neutrino Monte Carlo (MC) events generators.

Measurements of $M_A$ can use an information contained in the shape of the
distribution of events in the four-momentum transfer $Q^2$ (strictly speaking in the variable $Q^2_{QE}$, see \cite{MB_MA}) which is sensitive 
enough for precise evaluations of $M_A$. 
The dependence of the total cross-section on $M_A$ gives an additional input:
if the $M_A$ value is increased from $1.03$ 
to $1.35$~GeV for $E_\nu>1$~GeV the cross-section is raised by 
$\sim 30\%$ (for $E_\nu<1$~GeV the increase is smaller). 
Another interesting option to validate models is to compare to high statistics double differential (2D) cross section data (muon 
kinetic energy and scattering angle) on carbon provided by the MB collaboration \cite{MB_MA}.

In the past, several measurements of $M_A$  were done on 
deuterium for which most of nuclear physics complications 
are absent. Until a few years ago it seemed that the results 
converge to a value $\sim 1.03$~GeV \cite{bodek_MA}. There is an additional 
argument in favor of a similar value of $M_A$ coming from the weak 
pion-production at low $Q^2$. 
When put together they suggest the value $M_A=1.014$~GeV \cite{bodek_MA}.
On the contrary, all (with an exception 
of the NOMAD experiment) more recent measurements of 
$M_A$ report much larger values (for a discussion see: \cite{CCQE_sitges}). 

A theoretical mechanism which can explain the $M_A$ value discrepancy 
comes from the many-body nuclear model proposed
10 years ago \cite{marteau_suma} based on the ideas of M. Ericson and developed more recently by 
Martini, Ericson, Chanfray and Marteau (MEChM model).  
The model predicts a large contribution to the muon inclusive CC cross section from elementary 
2p-2h and 3p-3h excitations leading to multinucleon ejection. 
The contribution is 
absent in a free nucleon neutrino reaction and in the MB event selection is treated as CCQE giving rise to effective large $M_A$ value.

\begin{figure}
\resizebox{0.52\textwidth}{!}{%
  \includegraphics{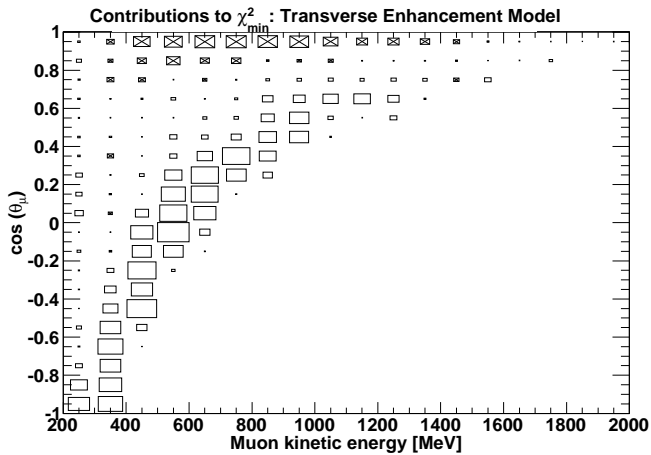}
}
\caption{TEM: contributions do $\chi^2_{min}$. Contributions in bins are proportional to the area of boxes. If a box is crossed the model prediction is larger then the experimental data}
\label{bodek_genuine}       
\end{figure}
A microscopic evaluation of the multinucleon 
ejection contribution was reported in \cite{nieves_suma}.
The computations were done in the theoretical scheme which was
succesfull in describing electron scattering in the kinematical region 
of QE and $\Delta$ peaks together with the 
{\it dip} region between them.
The model was applied to MB 2D cross section data 
and a fit to the axial mass value was done. 
In the fitting procedure 
\cite{agostini}  the authors included an 
overall $10.7\%$  normalization error.
The two-parameter fit gave results: $M_A=1.077\pm 0.027$~GeV and
for the normalization scale: $\lambda = 0.917\pm 0.029$. Using 
the low-momentum cut procedure, as proposed in \cite{jsz}, with 
$q_{cut}=400$~MeV the value $M_A=1.007\pm
0.034$~GeV was obtained.
\begin{figure}
\resizebox{0.52\textwidth}{!}{%
  \includegraphics{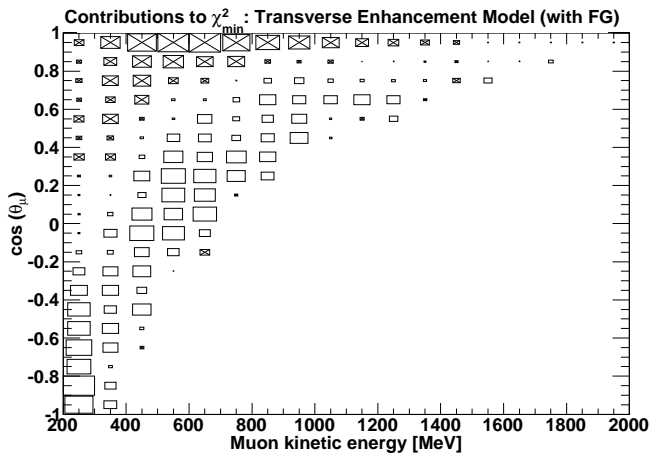}
}
\caption{TEM-FG: contributions do $\chi^2_{min}$. }
\label{bodek_FG}       
\end{figure}
\begin{figure}
\resizebox{0.52\textwidth}{!}{%
  \includegraphics{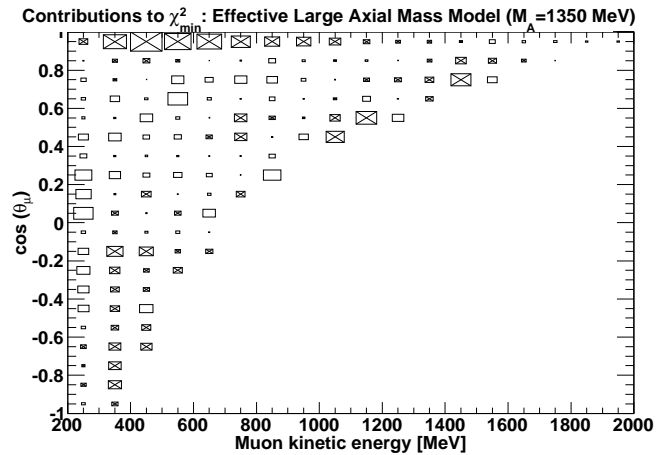}
}
\caption{ELAMM: contributions do $\chi^2_{min}$}
\label{FG}       
\end{figure}
There is still another approach to include 2p-2h contribution coming from Meson Exchange Current (MEC) diagrams \cite{scaling_2p2h}. It is shown that with the MEC contribution one gets closer to the MB experimental results.

\section{Transverse Enhancement Model (TEM)}
\label{sec:2}

In \cite{bodek} a new approach to model CCQE scattering on nuclear targets was proposed. The approach is intended to be easy to implement in MC event generators. It assumes that it is sufficient to describe properly an enhancement of the transverse electron QE response function keeping all other ingredients as in the free nucleon target case. 

The authors of \cite{bodek} proposed a transverse enhancement function for the carbon target. For low $Q^2$ its form is determined by the scaling arguments while for high $Q^2$ ($>0.5$~GeV$^2$) it is obtained as a fit to the inclusive electron cross section data from the JUPITER experiment. 
The prescription to include TE contribution in the numerical computations amounts to rescaling of the magnetic proton and neutron form factors:
\begin{equation}
G_M^{p,n}(Q^2)\rightarrow \sqrt{ 1 + AQ^2 \exp (-\frac{Q^2}{B}) } G_M^{p,n}(Q^2)
\end{equation}
where $A= 6$~GeV$^{-2}$ and $B=0.34~GeV^{2}$.

TEM model offers a chance to explain an apparent contradiction between recent low (MB) and high (NOMAD) neutrino energy $M_A$ measurements: for energies up to $\sim 700$~MeV the model predicts the CCQE cross section similar to effective large axial mass predictions with $M_A=1.3$~GeV. For higher neutrino energies the TEM cross section becomes smaller and at $E_\nu \sim 5$~GeV corresponds to $M_A\sim 1.15$~GeV.

The aim of this paper is to confront the predictions of the TEM model with the MB CCQE data. 
\begin{figure*}
\resizebox{1.0\textwidth}{!}{%
  \includegraphics{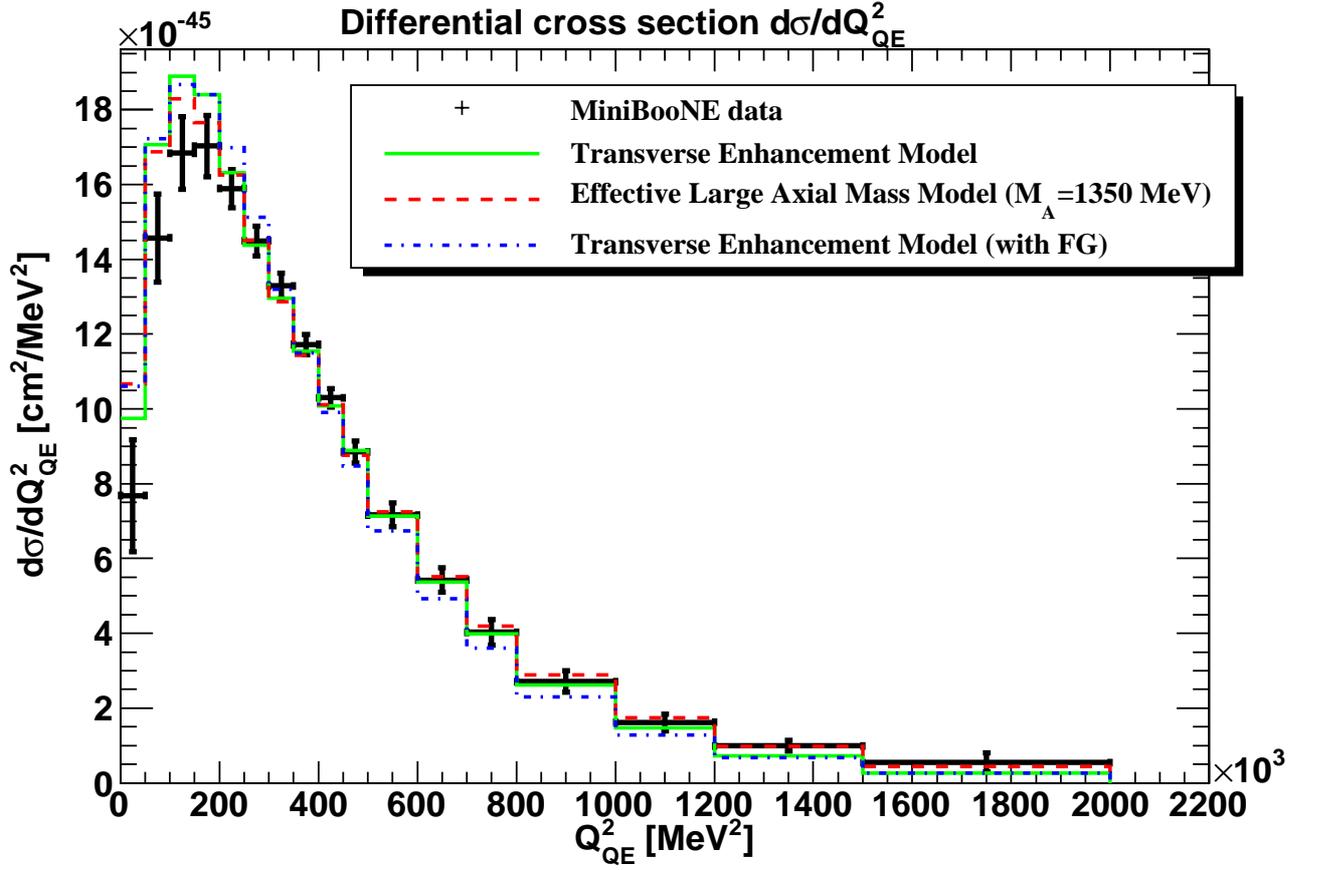}
}
\caption{ Differential cross section in $Q^2$}
\label{Q2}       
\end{figure*}
\section{Results and discussion}
\label{sec:3}

In our numerical analysis we compare predictions from two models: 

A) effective large axial mass model (ELAMM) wih $M_A=1.35$~GeV together with the FG model with parameter values as in the MB experimental analysis: $p_F=220$~MeV and $B=34$~MeV. 

B) TEM with the standard axial mass $M_A=1.014$~GeV (as used in \cite{bodek}). We investigate two implementations of the TEM: (i)
as in the original paper: without Fermi motion and with Pauli blocking effect introduced by means of the NEUGEN $Q^2$ dependent reduction function; (ii) with the Fermi motion and Pauli blocking implemented via the FG model; we call the model: TEM-FG.

In both models modifications of the standard ($M_A\sim 1.03$~GeV) theory are introduced in the $Q^2$ dependent way in agreement with the MB analysis of the 2D distribution of final muons (see Figs 11,12 in \cite{MB_MA}).

We produced three samples of $10^6$ events using NuWro MC event generator \cite{nuwro}. We checked that statistical fluctuations are small. Because in the MB data there is a large overall flux (normalization) error we introduce a renormalization factor to the $\chi^2$ statistical test defined as \cite{agostini}:
\begin{equation}\label{chi2} 
\chi^2_{} (\lambda) = \left(\frac{\lambda^{-1} -1}{ \Delta\lambda }\right)^2 +
\end{equation}
\[\sum_{i=1}^{137}
\left(\frac{\left( \frac{d^2\sigma } {dT_\mu d\cos\theta }\right)^{exp}_j  - \lambda \left( \frac {d^2\sigma } {dT_\mu d\cos\theta } \right)^{th}_j }{  \Delta \left( \frac { d^2\sigma} { dT_\mu d\cos\theta} \right)_j } \right)^2 \]
with $\Delta\lambda = 0.107$. It means that basically we compare the shapes of two-dimensional distributions of events.

\begin{table}
\caption{Results for 2D and $Q^2$ fits}
\label{tabela}       
\begin{tabular}{l || ll | ll}
\hline\noalign{\smallskip}
model & $\lambda_{2D}$ & $\chi^2_{2D,min}$ &$\lambda_{Q^2}$ &  $\chi^2_{Q^2,min}$\\
\noalign{\smallskip}\hline\noalign{\smallskip}
ELAMM & 1.03 & 34.1 & 1.075 & 15.8\\
TEM & 1.03 & 196.2 & 1.015 & 22.3 \\
TEM-FG & 1.135 & 133.3 & 1.08 & 44.0 \\
\noalign{\smallskip}\hline
\end{tabular}
\end{table}
Results are shown in Table \ref{tabela} in the second and third columns. The number of degrees of freedom is $DOF= 137$ (the number of non-zero bins) -$1=136$. 
We see that both $\chi^2_{TEM, min}$ and $\chi^2_{TEM-FG, min}$ are larger then $\chi^2_{ELAMM, min}$. Additionally, the value $\frac{\chi^2_{ELAMM,min}}{DOF}=\frac{34.1}{136}\approx 0.25$ is much smaller then $1$ which suggest that the shape errors were evaluated in a too conservative way.

Figs \ref{bodek_genuine}, \ref{bodek_FG}  and \ref{FG} show contributions to $\chi^2_{min}$ from three models. Contributions in bins are proportional to the area of boxes. If a box is crossed the model prediction is larger then the experimental data. We see that three patterns are rather different in shape and in the case of Figs. \ref{bodek_genuine} and \ref{bodek_FG} they may indicate that the models do not reproduce the cross section $Q^2$ dependence very well. We will come back to this point later. In all the cases there is a significant deficit of events in the region $\cos\theta_\mu\sim 1$. These are low $Q^2$ events for which it is known that techniques going beyond the FG (like RPA or CRPA) should be used \cite{rpa}. 

We made a similar statistical analysis with the $Q^2_{QE}$ differential cross section data. The number of bins is smaller $(17)$ but the uncorrelated relative shape errors are also much smaller. The $\chi^2$ is defined as:
\begin{equation}
\label{chi2_Q2} \chi^2_{} (\lambda) =\left(\frac{\lambda^{-1} -1}{ \Delta\lambda }\right)^2 +
\end{equation}
\[\sum_{i=1}^{17}
\left(\frac{\left( \frac{d\sigma } {dQ^2_{QE}}\right)^{exp}_j  - \lambda \left( \frac {d\sigma } {dQ^2_{QE}} \right)^{th}_j }{\Delta \left( \frac { d\sigma} {dQ^2_{QE}} \right)_j } \right)^2.\]
Results are shown in Table \ref{tabela} in the last two columns. 
The number of degrees of freedom is now $DOF= 17$ (the number of bins) -$1=16$. The results indicate that also here the ELAMM performance is better then that of TEM and TEM-FG. Fig \ref{Q2} shows the MB data and three models predictions at the best fit points. In all the models there is a significant disagreement with the data at $Q^2_{QE}<0.1$~GeV$^2$. At large $Q^2_{QE}$ the ELAMM cross section is bigger then that of TEM and TEM-FG in agreement with Figs. 7 and 8 from \cite{bodek}.

\begin{figure}
\resizebox{0.52\textwidth}{!}{%
  \includegraphics{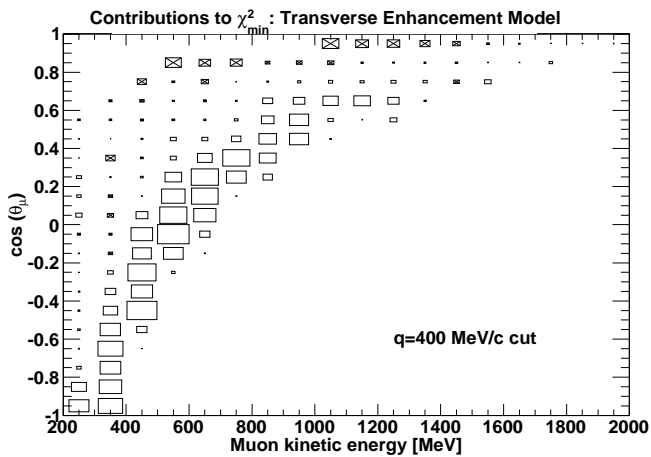}
}
\caption{TEM: contributions do $\chi^2_{min}$ with the low momentum cut $q_{cut}=400$~MeV/c.}
\label{bodek_genuine_cut}       
\end{figure}

\begin{figure}
\resizebox{0.52\textwidth}{!}{%
  \includegraphics{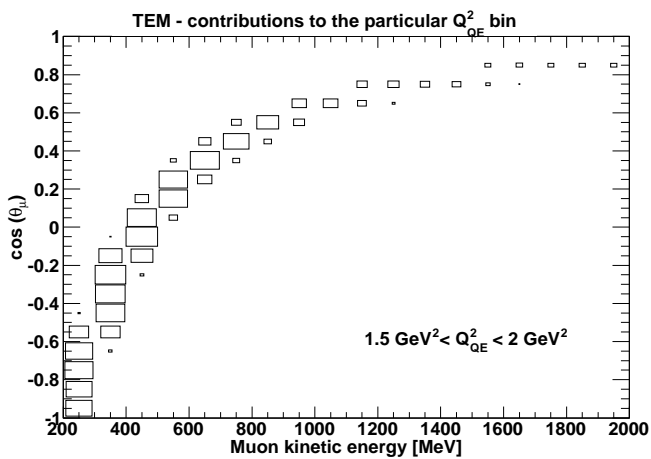}
}
\caption{Contributions to the $Q^2_{QE}\in (1.5, 2)$~GeV$^2$ bin}
\label{vivi}       
\end{figure}

We investigated an impact of the low $Q^2$ bins on the final results and aplied the low momentum transfer cut $q_{cut}=400$~MeV/c, as explained in \cite{jsz}. 2D best fit values of $\chi^2$ were reduced to $170.8$, $98.7$ and $23.3$ for TEH, TEH-FG and ELAMM respectively and corresponding best fit values of $\lambda$ were only slightly increased (by about 2-3\%). Fig. \ref{bodek_genuine_cut} shows the contributions to $\chi^2_{TEM, min}$ with the low momentum cut applied. 

Finally, we tried to understand why for the TEM model $\chi^2_{Q^2, min}$ is relatively small even if there are 2D bins producing very large contribution to $\chi^2_{2D, min}$. We selected three such bins: $T_\mu\in (500,600)$~MeV and 
$\cos\theta_\mu\in (-0.1, 0.2)$ with contributions: 9.5, 7.2 and 5.5 
(see: \ref{bodek_genuine_cut}). We checked that according to TEM events from the selected bins contribute only to two
$Q^2_{QE}$ bins: $(1.2, 1.5)$~GeV$^2$ and 
$(1.5, 2)$~ GeV$^2$ (cross sections
$5.42\cdot 10^{-48}$~cm$^2$ and 
$1.44\cdot 10^{-47}$~cm$^2$/MeV$^2$ respectively). 
Fig \ref{vivi} shows that these contributions represent a small fraction of the overall cross section in the last $Q^2_{QE}$ bin and the large disagreement in three selected bins is hidden in the averaged $Q^2_{QE}$ analysis. It is clear that in order to get a deep insight into CCQE, the complete 2D data should be analyzed. 

We conlude that it seems that the effective large axial mass model leads to better agteement with the MB data. However, one should remember that a real challenge is to provide predictions for the hadronic final states in multinucleon ejection effectively described by either ELAMM or TEM.

\vskip 0.25cm
{\bf Acknowledgments:}
The author was supported by the grants: N N202 368439 and DWM/57/T2K/2007.

\end{document}